\documentclass[12pt, amssymb, nofootinbib, aps, prd]{revtex4}

\makeatletter\AtBeginDocument{\let\@elt\relax}\makeatother 

\usepackage[english]{babel}
\usepackage{amsmath}
\usepackage{amssymb}
\usepackage{amsbsy}
\usepackage{amstext}
\usepackage{graphicx}
\usepackage{float}
\makeatletter\AtBeginDocument{\let\@elt\relax}\makeatother
\usepackage{hyperref}
\usepackage{pst-node}
\usepackage{verbatim}
\usepackage{tikz}
\usepackage{cancel}
\usepackage{braket}
\usepackage[normalem]{ulem}
\usetikzlibrary{snakes,decorations.pathmorphing}

\newcommand{\be}{\begin{eqnarray}}
\newcommand{\ee}{\end{eqnarray}}
\newcommand{\bdm}{\begin{displaymath}}
\newcommand{\edm}{\end{displaymath}}
\newcommand{\ds}{\displaystyle}
\newcommand{\nn}{\nonumber}
\newcommand{\ba}{\begin{array}}
\newcommand{\ea}{\end{array}}
\newcommand{\pa}[1]{\left(#1\right)}
\newcommand{\paq}[1]{\left[#1\right]}

\newcommand{\K}{{\bf k}}
\newcommand{\Q}{{\bf q}}

\newcommand{\X}{{\bf x}}

\newcommand{\eps}{\epsilon}

\begin{document}

\title{Conservative binary dynamics\\
 from gravitational tail emission processes}

\author{Gabriel Luz Almeida}
\email{gabriel.luz@fisica.ufrn.br}
\affiliation{Departamento de F\'\i sica Te\'orica e Experimental, Universidade Federal do Rio Grande do Norte, Avenida Senador Salgado Filho, Natal-RN 59078-970, Brazil}

\author{Alan M\"{u}ller}
\email{alan.muller@unesp.br}
\affiliation{Instituto de F\'\i sica Te\'orica, UNESP - Universidade Estadual Paulista, S\~ao Paulo 01140-070, SP, Brazil}

\author{Stefano Foffa}
\email{stefano.foffa@unige.ch}
\affiliation{D\'epartement de Physique Th\'eorique and Gravitational Wave Science Center, Universit\'e de Gen\`eve, CH-1211 Geneva, Switzerland}

\author{Riccardo Sturani}
\email{riccardo.sturani@unesp.br}
\affiliation{Instituto de F\'\i sica Te\'orica, UNESP - Universidade Estadual Paulista \& ICTP South American Institute for Fundamental Research, S\~ao Paulo 01140-070, SP, Brazil}

\begin{abstract}
  We re-analyze the far zone contribution to the two-body conservative dynamics
  arising from interaction between radiative and longitudinal modes, the latter
  sourced by mass and angular momentum, which in the mass
  case is known as tail process.
  We verify the expected correspondence between two loop self-energy amplitudes
  and the gluing of two classical (one leading order, one at one loop) emission
  amplitudes, with focus on the Ward identities.
As part of our analysis, we originally compute emission and self-energy processes with the longitudinal mode sourced by angular momentum for generic electric and magnetic multipoles
and we highlight the role of the contribution from source interaction with
  two gravitational fields.
\end{abstract}

\maketitle

\section{Introduction}

The recent advent of Gravitational Wave (GW) astronomy opened a new field
not only for observing the universe, but also for investigating the fundamental
nature of gravity in the most profound way so far possible.

GW detections \cite{LIGOScientific:2021djp} by the interferometric LIGO
\cite{TheLIGOScientific:2014jea} and Virgo \cite{TheVirgo:2014hva} observatories
have collected signals from compact coalescing binaries in the three completed science
runs, with the fourth one presently ongoing.
Moreover, a third generation of terrestrial detectors and a space detector are
already planned for the next decade, expecting to reach
signal-to-noise ratios at $O(10^3)$
\cite{Maggiore:2019uih,Reitze:2019iox,LISA:2022kgy}. 

Besides representing a new way to observe (or rather \emph{listen to}) the cosmos,
such GW signals are privileged windows to investigate the nature of gravity
at unprecedented strongly interacting level.
Detections require correlation of data with pre-computed waveforms \cite{mf,LIGOScientific:2019hgc},
whose accuracy is crucial in maximizing the physics output of observations,
and which, in turn, depends on precise knowledge of the two-body dynamics.

In view of deepening our analytic insight into the two body dynamics, we
(re-)investigate in the present work the processes arising from the scattering
of radiation off the static curvature produced by the same sources of radiation.
Such processes, besides representing corrections to the emission process
at one (classical) loop, also play a role in two-loop self-energy diagrams contributing to the binary dynamics.
In particular the scattering of radiation off the static curvature sourced
by the mass of the system is known as \emph{tail} process \cite{Blanchet:1987wq} ($M$-tail in this paper),
because of its phenomenological property of travelling \emph{inside} the light
cone, rather than on it, giving rise to a metaphorical ``tail strike''
in the emitted wave.
Tails represent one class of \emph{hereditary} processes, i.e., processes
relating the field at the observer to the entire history of
the source, rather than its instantaneous state at retarded time.
Another class of hereditary processes is represented by \emph{memory} ones
\cite{Christodoulou:1991cr},
where radiation scatters onto itself, and whose investigation we reserve to
a successive study.

The scattering of radiation off the static curvature generated by the angular
momentum can be described in complete analogy with the tail at the fundamental
level, but
its phenomenological effect is different as it gives rise to an instantaneous
effect in the waveform, hence it has been dubbed \emph{failed} tail \cite{Foffa:2019eeb}, which will be referred to as \emph{$L$-ftail} henceforth.

Within conservative dynamics, separation of processes into those involving the
exchange of potential modes only, and those involving also radiative modes,
is a standard procedure of perturbative computations.
  This separation is realized by
the \emph{method of regions} applied to particle physics \cite{Beneke:1997zp,Smirnov:1994tg} and the standard
near/far zone distinction in traditional gravitational computations \cite{Blanchet:2013haa}.

Restricting here to the spin-less case, among the perturbative
approaches used to investigate the conservative dynamics, we highlight the post-Minkowskian (PM) and post-Newtonian
(PN) approximation
schemes, which are best suited for studying respectively unbound scatterings
and bound systems.
The expansion parameters of the former is $GM/b$, being $G$ Newton's constant, $M$ the total mass of the two-body system, $b$ the impact parameter,
and for the latter $v^2\sim GM/r$, being $v$ the relative velocity of
binary constituents and $r$ the size of the orbit.

In the context of PM approximation, dynamics has been completed at 4PM level \cite{Bern:2021dqo,Bern:2021yeh,Dlapa:2021npj,Dlapa:2021vgp},
considering the standard \emph{Feynman} prescription for Green functions of
radiative internal modes, which is consistent with the \emph{principal value}
prescription, corresponding to time-symmetric Green functions, adopted in
\cite{Damour:2016gwp}. Note that, as far as conservative dynamics is
  concerned, for self-energy processes involving \emph{up to two} internal
  radiative Green functions it is indeed equivalent to using Feynman or
  retarded/advanced Green functions. However a treatment in terms of the in-in formalism
\cite{Galley:2009px} is necessary for the memory processes \cite{Blumlein:2021txe,Almeida:2021xwn,Almeida:2022jrv}.

Results obtained with amplitude or effective field theory (EFT) methods within
the PM formalism, see also \cite{Dlapa:2022lmu,Kalin:2022hph}, can be framed in
an
elegant and compact form in terms of two-body scattering angle $\chi$,
whose PM perturbative expansion has a simple and distinctive
scaling with the symmetric mass ratio $\eta\equiv m_1m_2/M^2$, being
$m_{1,2}$ the individual binary constituent masses.
It has been shown in \cite{Damour:2019lcq} that the $m$-PM contribution to the scattering angle, $\chi_m$, scales with $\eta$ as
$\chi_m\sim \eta^{[(m-1)/2]}$.
In particular \cite{Kalin:2022hph} included non-time symmetric radiation reaction effects, but still does not return the correct scaling of $\chi_4$ with
$\eta$.

Far zone processes in conservative two-body dynamics have been already considered
by several studies, see e.g.~\cite{Foffa:2019eeb,Blumlein:2021txe,Almeida:2021xwn,Almeida:2022jrv}, up to 5PN level, i.e.~considering tail, $L$-ftail \emph{and}
memory effects, but their result is still inconsistent with the above-mentioned $\eta$
scaling of $\chi_4$, requiring further work to solve the discrepancy
\cite{Bini:2021gat,Bini:2022enm}.

Here, we investigate how tail processes can be analysed in terms
of \emph{generalized unitarity}
\cite{Bern:1994cg,Bern:1994zx}, i.e. how self-energy diagrams can be described
by gluing a pair of emission diagrams in the context 
of the Non-Relativistic General Relativity (NRGR) EFT approach
\cite{Goldberger:2004jt} to the gravitational two-body problem within the PN
approximating scheme. See \cite{Edison:2022cdu,Edison:2023qvg} for an
application to generalized unitarity use in NRGR.

With the goal of shedding light on the EFT side of the problem, we revisited
the study of processes involving $M$-tail and $L$-ftail, showing that the
computation procedure adopted so far \cite{Foffa:2019eeb,Blumlein:2021txe,Brunello:2022zui} leads to an incorrect result for the $L$-ftail.
In particular in the case of the electric quadrupole this is due to having
  overlooked the process involving a source-graviton-graviton interaction
  vertex (henceforth quadratic-interaction), negligence leading to
 an apparent violation of the Ward identities, or equivalently a violation to the
gauge fixing condition.
In the case of the magnetic quadrupole, the Ward identities
can be restored by adding to the action a local term altering
the equations of motion and leading to solutions satisfying the gauge condition
and the energy-momentum conservation, along a procedure first used in the
multipolar PM approach in \cite{Blanchet:1997ji}.
For higher order multipoles, the gauge condition is automatically satisfied.

Our new result confirms the independent result of \cite{Henry:2023sdy}
and the known computation of the angular momentum flux
ascribable to the $L$-ftail process in \cite{Arun:2009mc}.

The paper is structured as follows. In section \ref{sec:intro} we set up notations and write down emission amplitudes for generic  multipoles, first at the leading order, then at next-to-leading in $G$ for $M$-tails and $L$-ftails, the latter being derived here for the first time for generic
  electric and magnetic multipoles.
We show that the quadratic-interaction process gives a critical contribution to the $L$-ftail involving the electric quadrupole. The section is ended by a discussion of a residual 
violation of the Lorentz gauge, happening in the magnetic quadrupole case.
In section \ref{sec:cuts} we investigate the relation between self-energy
diagrams and (square of) emission processes, and how the former are impacted by
the Ward identities issue affecting the $L$-ftail.
We use angular momentum balance equation to confirm our new value for the
$L$-ftail self-energy diagram involving the electric quadrupole, and derive for
the first time the corresponding values for all electric and magnetic multipoles.
Section \ref{sec:concl} contains our conclusions and prospects, while some technical derivations are detailed in the appendices.

\section{Emission amplitudes in EFT}
\label{sec:intro}

\subsection{One point functions and leading order emission amplitudes}
Our starting point is the following multipolar action which gives the linear
coupling of matter to the gravitational field:
\be
\label{eq:ssource}
S_{\rm source} &=& \int_t\,\left[ \frac12 E h_{00} -\frac12 J^{k|l} h_{0k,l} - \sum_{r\ge 0}\left( c_r^{(I)} I^{ijR} \partial_R {\cal R}_{0i0j} + \frac{c_r^{(J)}}{2} J^{k|iRl} \partial_R {\cal R}_{0ilk} \right) \right]\,,
\ee
with
\be
c_r^{(I)} = \frac{1}{(r+2)!}\,, \qquad c_r^{(J)} = \frac{2(r+2)}{(r+3)!}\,,
\ee
and where $J^{k|iRl}$ are the $d$-dimensional generalizations \cite{Henry:2021cek} of the 3-dimensional magnetic-type multipoles $J^{ijR}=\frac12 \epsilon_{kl(i} \left.J^{k|jR)l}\right|_{d=3}$ (symmetrized over the indices $ijR$) and $I^{ijR}$ the standard electric-type ones.\footnote{We adopted the notation $\int_x\equiv \int{\rm d}x$, $\int_\K\equiv \int \frac{{\rm d}^{d}k}{\pa{2\pi}^d}$. Our metric convention is ``mostly plus'':
  $\eta_{\mu\nu}={\rm diag}\pa{-1,+1,+1,+1}$. }
In expression (\ref{eq:ssource}), ${\cal R}_{\mu\nu\rho\sigma}$ are the components
of the Riemann curvature tensor, while $R$ denotes the collective symmetric
trace-free index, $R=i_1\dots i_r$, with $r=0$ standing for the quadrupole, $r=1$ for the
octupole, and so on.

The dynamics of the gravitational field is dictated by the bulk action, given by the Einstein-Hilbert action plus a gauge-fixing term:
\begin{align}\label{bulkgravity}
    S_{\rm bulk} &= 2\Lambda^2 \int {\rm d}^{d+1}x \sqrt{-g} \left[ {\cal R}(g) - \frac12 \Gamma_\mu \Gamma^\mu \right]\,,
\end{align}
with $\Lambda\equiv \pa{32 \pi G_d}^{-1/2}$, with $G_d$ being Newton's constant in $d+1$-dimensions,
$\Gamma^\mu\equiv\Gamma^\mu_{\nu\rho}g^{\nu\rho}$, being $\Gamma^\mu_{\nu\rho}$ the
standard Christoffel connection, and the metric $g_{\mu\nu}$ will be eventually
expanded around Minkowski background as per $g_{\mu\nu}=\eta_{\mu\nu}+h_{\mu\nu}$.

The gauge-fixing term implies that the theory we are solving for falls back
into GR only for $\Gamma^\mu=0$, which implies the Lorentz
condition $\partial^\nu \bar h_{\mu\nu}=0$ at linearized level.
An overbar denotes the
trace-reversed field $\bar h_{\mu\nu}\equiv h_{\mu\nu}-\frac 12 \eta_{\mu\nu}h$, being
$h\equiv \eta^{\mu\nu}h_{\mu\nu}$.
The addition to the Lagrangian of the gauge-fixing term modifies the linearized
Einstein equations precisely to $\Box \bar{h}_{\mu\nu}=0$ outside the source.

From the quadratic term in Eq.~\eqref{bulkgravity}
 we can derive the Green's functions for the gravitational perturbation field $h_{\mu\nu}\equiv g_{\mu\nu}-\eta_{\mu\nu}$:
\begin{equation}
\label{eq:propagator}
P[h_{\mu\nu},h^{\alpha\beta}]=-\frac{i}{\K^2-\omega^2}\frac{{{\cal P}_{\mu\nu}}^{\alpha\beta}}{\Lambda^2}\,,\quad {{\cal P}_{\mu\nu}}^{\alpha\beta}\equiv \frac 12 \pa{\delta_\mu^\alpha\delta_\nu^\beta+\delta_\mu^\beta\delta_\nu^\alpha-\frac 2{d-1}\eta_{\mu\nu}\eta^{\alpha\beta}}\,,
\end{equation}
where so far we have not specified the Green's function boundary conditions.
GWs correspond to the transverse traceless spatial component of the metric
perturbations, and for direction propagation ${\bf \hat n}$ they can be selected by applying the following projector operator
\be
\Lambda^{TT}_{ij,kl}({\bf \hat n})\equiv P_{ik}P_{jl}-\frac1{d-1} P_{ij}P_{kl}\,,\quad P_{ij}({\bf \hat n})\equiv\delta_{ij}-\hat n_i \hat n_j\,.
\ee
Beside gauge ones, the other components parametrize longitudinal degrees of freedom, which
play a role for checking energy-momentum conservation, as it will be discussed
below.

We denote by $i\mathcal{A}_{\alpha\beta}(\omega,\K){h^*}^{\alpha\beta}(\omega,\K)$
the probability amplitude for the emission of the generic
field $h_{\alpha\beta}$, which can be computed by deriving the appropriate Feynman rules from \eqref{eq:ssource} and \eqref{bulkgravity}.
The classical field at a spacetime position $x$,
given by the one-point function $\braket{h_{\mu\nu}(x)}$, is then related to $\mathcal{A}_{\mu\nu}$ by:
\begin{equation}
\label{eq:hexpect}
\braket{h_{\mu\nu}(x)} = \int\mathcal{D}h\, e^{iS[h]}h_{\mu\nu}(x)=\int_\K \frac{{\rm d}\omega}{2\pi} \frac{e^{-i\omega t+i\K\cdot\X}}{\K^2-(\omega+i{\tt a})^2}
\frac{{{\cal P}_{\mu\nu}}^{\alpha\beta}}{\Lambda^2}{\cal A}_{\alpha\beta}(\omega,\K)\,,
\end{equation}
where the correct retarded boundary condition has been selected.\footnote{
We displace the Green's function pole by $\pm i{\tt a}$, as $\eps$ is already
used to denote $d-3$.}

The linearized Lorentz gauge condition translates into ``Ward'' identities
for the classical process consisting of the emission of a single gravitational
mode\footnote{Note that the trace reversion operator ${{\hat{\cal P}}_{\mu\nu}}^{\alpha\beta}\equiv \frac 12 \pa{\delta_\mu^\alpha\delta_\nu^\beta+\delta_\mu^\beta\delta_\nu^\alpha-\eta_{\mu\nu}\eta^{\alpha\beta}}$, which turns $h_{\alpha\beta}$ into $\bar{h}_{\mu\nu}$, is identical to its inverse ${{\cal P}_{\mu\nu}}^{\alpha\beta}$ only for $d=3$.}

\begin{equation}\label{eq:Lorentz}
  \partial^\mu\braket{\bar{h}_{\mu\nu}(x)} = 0 \quad \Leftrightarrow \quad
  k^\mu{\cal A}_{\mu\nu}(\omega,\K) = 0\,.
\end{equation}

As it happens in the multipolar PM approach \cite{Blanchet:2013haa}, the
diagrammatic expansion used in our EFT setup provides a solution of the
perturbative form of Einstein's equations
\begin{equation}
\Box \bar{h}_{\mu\nu} = \Lambda_{\mu\nu}\,,
\end{equation}
where $\Lambda_{\mu\nu}$ is a source term given by the source energy-momentum
tensor plus non linear terms in $h_{\mu\nu}$.
The gauge condition (\ref{eq:Lorentz}), which is equivalent to the conservation
of the pseudo-energy momentum tensor $\Lambda_{\mu\nu}$, is however not automatically
satisfied and it has to be checked, and \emph{eventually fixed},
on a case-by-case basis.

It is not uncommon in field theory that loop interactions can break a symmetry
which is present in the free theory, causing the symmetry to be \emph{anomalous}.
In our case GR invariance under diffeomorphism is broken in the Lagrangian
by the gauge fixing term, implying that GR is recovered only when $\Gamma^\mu=0$,
which makes the gauge-fixing term a \emph{double zero}. We will see in
sec.~\ref{ssec:tails} that relation (\ref{eq:Lorentz}) may not hold at
interacting level, but in a \emph{consistent}, or \emph{integrable} way,
i.e.~it is possible to add to the action functional a local term restoring the
Lorentz condition (\ref{eq:Lorentz}) at the level of the equations of motion.

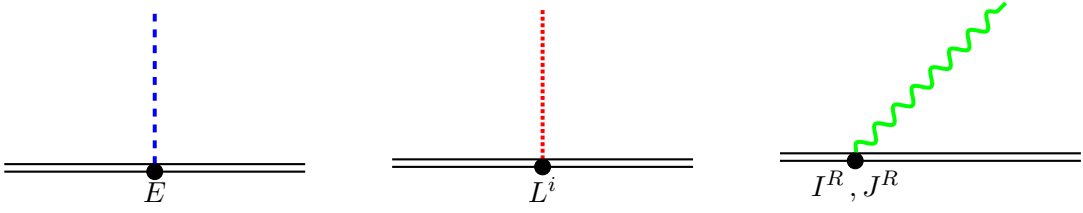
\begin{figure}
	\begin{center}
		\begin{tikzpicture}
      			\draw [black, thick] (0,0) -- (4.,0);
      			\draw [black, thick] (0,-0.1) -- (4.,-0.1);
      			\draw [blue, dashed, line width=1.5] (2,0) -- (2,2);
      			\filldraw[black] (2.,-0.1) circle (3pt) node[anchor=north] {$E$};
		\end{tikzpicture}\hspace{1cm}
		\begin{tikzpicture}
      			\draw [black, thick] (0,0) -- (4.,0);
      			\draw [black, thick] (0,-0.1) -- (4.,-0.1);
      			\draw [red, densely dotted, line width=1.5] (2,0) -- (2,2);
      			\filldraw[black] (2.,-0.1) circle (3pt) node[anchor=north] {$L^i$};
		\end{tikzpicture}\hspace{1cm}
		\begin{tikzpicture}
      			\draw [black, thick] (0,0) -- (4.,0);
      			\draw [black, thick] (0,-0.1) -- (4.,-0.1);
    			\draw[decorate, decoration=snake, line width=1.5pt, green] (1,0) -- (3,2) ;
      			\filldraw[black] (1.,-0.1) circle (3pt) node[anchor=north] {$I^R\,, J^R$};
		\end{tikzpicture}\hspace{1cm}
	\end{center}
	\caption{Leading order emission amplitudes. Blue dashed line refers
          to a $h_{00}$ polarization, dotted red to $h_{0i}$, wavy green to
          $h_{ij}$.}
	\label{fig:simple}
\end{figure}

The conserved multipoles (mass and angular momentum) of action (\ref{eq:ssource}),
see the first two diagrams of Fig.~\ref{fig:simple}, contribute to
(\ref{eq:hexpect}) as:
\be
\ba{rcl}
\ds\mathcal{A}^{(c)}_{00}(\omega,\K)&=&\ds \frac12 E(\omega)\,,\\
\ds\mathcal{A}^{(c)}_{0k}(\omega,\K)&=&\ds \frac 14i k_j\epsilon_{ijk}L_i(\omega)\,,\\
\ds\mathcal{A}^{(c)}_{kl}(\omega,\K)&=&\ds 0\,,
\ea
\ee
where the finiteness of the amplitude for $d=3$ allowed us to use the
standard expression of the $L_{i}=\frac12\epsilon_{ijk}J^{j|k}$.
The Ward identities follow from
\be
\omega E(\omega)=0\,,\quad \omega L_i(\omega)=0\,,
\ee
and are trivially satisfied at this perturbative order by admitting that $E$
and $L_i$ are conserved.

The leading order electric and magnetic multipole emission amplitudes,
see last diagram in Fig.~\ref{fig:simple}, are finite for $d=3$ and read
\be
\left.\begin{array}{c}
i \mathcal{A}^{(e)}_{00}(\omega,\K)\\
i \mathcal{A}^{(e)}_{0k}(\omega,\K)\\
i \mathcal{A}^{(e)}_{kl}(\omega,\K)
\end{array}\right\}
= \frac12 c_r^{(I)}  (-i)^{r+1}k_R I^{ij R}(\omega)\times
\left\{\begin{array}{c}
k_i k_j \\
- \omega k_j \delta_{ik}\\
\omega^2 \delta_{ik} \delta_{jl}
\end{array}\right.\,,
\ee
and
\be
\left.\begin{array}{c}
i \mathcal{A}^{(m)}_{00}(\omega,\K)\\
i \mathcal{A}^{(m)}_{0k}(\omega,\K)\\
i \mathcal{A}^{(m)}_{kl}(\omega,\K)
\end{array}\right\}
=\frac12 c_r^{(J)}  (-i)^{r+1}  \epsilon_{imn} k_n  k_R J^{jmR}(\omega)\times
\left\{\begin{array}{c}
0 \\
-\frac12k_j \delta_{ik} \\
\omega \delta_{i(k} \delta_{l)j}
\end{array}\right.\,,
\ee
whose relative Ward identities are trivially satisfied.

\subsection{Tail-corrected emission amplitudes}
\label{ssec:tails}

\begin{figure}
	\begin{center}
		\begin{tikzpicture}
      			\draw [black, thick] (0,0) -- (4.,0);
      			\draw [black, thick] (0,-0.1) -- (4.,-0.1);
      			\draw [blue, dashed, line width=1.5] (2,0) -- (2,1.5);
			\draw[decorate, decoration=snake, line width=1.5pt, green] (.5,0) -- (2,1.5);
			\draw[decorate, decoration=snake, line width=1.5pt, green] (2,1.5) -- (3.5,2) ;
      			\filldraw[black] (2.,-0.1) circle (3pt) node[anchor=north] {$E$};
      			\filldraw[black] (.5,-0.1) circle (3pt) node[anchor=north] {$I^R\,, J^R$};
		\end{tikzpicture}\hspace{1cm}
		\begin{tikzpicture}
      			\draw [black, thick] (0,0) -- (4.,0);
      			\draw [black, thick] (0,-0.1) -- (4.,-0.1);
      			\draw [red, densely dotted, line width=1.5] (2,0) -- (2,1.5);
			\draw[decorate, decoration=snake, line width=1.5pt, green] (.5,0) -- (2,1.5);
			\draw[decorate, decoration=snake, line width=1.5pt, green] (2,1.5) -- (3.5,2) ;
      			\filldraw[black] (2.,-0.1) circle (3pt) node[anchor=north] {$L^i$};
      			\filldraw[black] (.5,-0.1) circle (3pt) node[anchor=north] {$I^R\,, J^R$};
		\end{tikzpicture}\hspace{1cm}
	\end{center}
	\caption{Next-to-leading order emission processes involving the
          scattering of radiation off the background static curvature sourced
        by energy and angular momentum.}
	\label{fig:tail}
\end{figure}
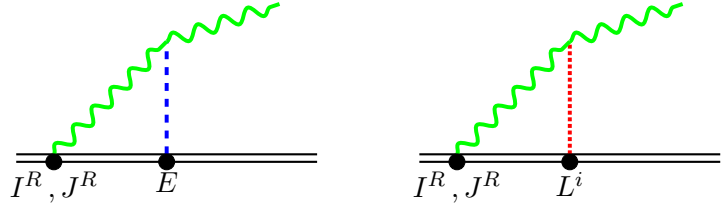

At next-to-leading order in gravitational interactions one has to consider
processes of the type represented in fig.~\ref{fig:tail}. The first one, involving
the scattering of radiation off the background curvature sourced by the system
total mass, gives a contribution to the waveform arriving later than the wavefront,
which propagates at the speed of light, while the tail propagates inside
the light cone.

The second process in Fig.~\ref{fig:tail}, involving the angular momentum, gives a purely local term
in the waveform, hence not giving rise to a \emph{tail}
effect. However its diagrammatic analogy with the tail process
suggests to lump it together with the $M$-tail, and we refer to it
in this work as $L$-ftail.

The $M$-tail amplitude is divergent for $d\rightarrow 3$ and its radiative,
transverse-traceless, on-shell ($\K=\omega {\bf \hat n}$)
part is \cite{Blanchet:1995fr,Almeida:2021jyt}
\be
\label{eq:Mtaile}
 \ds i \pa{\mathcal{A}^{(e-M{\rm -tail})}_{ij}}^{TT}(\omega,\omega {\bf \hat{n}}) &=&\ds (-i)^{r+1} \omega^2 c_r^{(I)} \left( i G E \omega \right) \Lambda^{TT}_{ij\,,kl}k_R I^{klR}(\omega) \times \left( \frac 1\epsilon - \kappa_{r+2}+\frac{\log x}2\right)\,,\\
\label{eq:Mtailm}
 \ds i \pa{\mathcal{A}^{(m-M{\rm -tail)}}_{ij}}^{TT}(\omega,\omega {\bf \hat{n}}) &=&\ds
 (-i)^{r+1} \omega c_r^{(J)}\left( i G E \omega\right) \Lambda^{TT}_{ij\,,kl} k_R k_nJ^{n|kRl}(\omega) \times \left( \frac 1\epsilon - \pi_{r+2} + \frac{\log x}2 \right)\,,\nonumber\\
 \\
\ds\kappa_{l} &=&\ds \frac{2l^2+5l+4}{l(l+1)(l+2)} + \sum_{i=1}^{l-2} \frac{1}{i}\,, \qquad  \pi_{l}=\frac{l-1}{l(l+1)} + \sum_{i=1}^{l-1} \frac{1}{i}\,,
\ee
where $\epsilon\equiv d-3$,
$x\equiv - e^{\gamma}\omega^2/\mu\pi$, and the inverse length scale $\mu$ is
implicitly defined by $G_d=G\mu^{-\epsilon}$, with $G$ denoting the standard $3+1$-dimensional Newton's constant.

The Ward identities
$k^\mu\mathcal{A}^{(e(m)-M{\rm -tail})}_{\mu\nu}(\omega,\omega{\bf \hat n}) = 0$
are conveniently found by taking the divergene of ${\cal A}_{\mu\nu}^{(e(m)-M-tail)}$
before performing the loop integration, as reported in Appendix
\ref{app:MLtailui}.

The $L$-ftail amplitudes are finite and local and the expressions of their
radiative TT part for generic electric and magnetic multipoles are originally given here
\be
\label{eq:eLftail_amp}
&&i\pa{\mathcal{A}^{(e-L{\rm -ftail})}_{ij}}^{TT}(\omega,\omega {\bf \hat{n}}) = (-i)^r c_r^{(I)}G \Lambda^{TT}_{ij\,,kl}\epsilon_{mnq}\frac{i\omega^2L^q I^{pR(k}(\omega)}{(r+1)(r+2)(r+3)(r+4)}\nonumber\\
&&\times\bigg\{ k_n \!\pa{2[6+r(r+4)]\delta_{l)m} k_p k_R \! - \! r(10+r(r+5))\delta_{l)p}\delta_{i_1 m}k_{R-1}\omega^2}\!+\! 24 \delta_{0r}\delta_{l)m}\delta_{np}k_R\omega^2\bigg\}\,,\\
\label{eq:mLftail_amp}
&&i\pa{\mathcal{A}^{(m-L{\rm -ftail})}_{ij}}^{TT}(\omega,\omega {\bf \hat{n}}) = - (-i)^r c_r^{(J)}G  \Lambda^{TT}_{ij\,,kl}\frac{i\omega^3 L^q J^{pR(k}(\omega)}{(r+1)(r+2)(r+3)(r+4)(r+5)}\\
&&\times\left\{\pa{r^4+10 r^3+35 r^2+50 r +48 }\delta_{l)p}\paq{k_q k_R-(1- \delta_{0r})\omega^2\delta_{qi_1}k_{R-1}}\right.\nn\\
&&\qquad\qquad\quad\qquad\qquad\qquad\qquad\qquad\qquad\qquad\left.-\pa{r^4+12 r^3+53 r^2+102 r+84}\delta_{l)q}k_p k_R\right\}\,.\nn
\ee
with their un-integrated form also reported in App. \ref{app:MLtailui}.

This time the spatial Ward identities are \emph{not} satisfied
\be
\label{eq:Weviol}
\ba{rcl}
\ds k^\mu\mathcal{A}^{(e-L{\rm -ftail})}_{\mu l}(\omega,\omega{\bf \hat n}) &=&\ds
(-i)^{r+1}\frac{c_r^{(I)}}{2\Lambda^2} \left( \frac{\omega}{4} \right)  
k_j \omega^2 \epsilon_{ijk} L^k I^{iRl}(\omega) \int_\Q \frac{q_R}{(\Q^2-\omega^2)}\\
&\overset{d\rightarrow 3}=&\ds
- \delta_{r0} \frac{G}2   k_j \omega^4 \epsilon_{ijk} L^k I^{il}(\omega)\,.
\ea
\ee
As the $\Q$-integral gives a result proportional to the symmetric, traceless
combination of Kronecker deltas with $R$ indices, expression (\ref{eq:Weviol})
vanishes except for $r=0$, meaning that the $L$-ftail emission process
involving the electric quadrupole violates the spatial components of the Ward
identity.
Analogously, for the magnetic part one finds 
\be
\label{eq:Wmviol}
k^\mu\mathcal{A}^{(m-L\rm-tail)}_{\mu l}(\omega,\omega{\bf \hat n}) &\overset{d\rightarrow 3}=&  \delta_{r0}\frac{4}{15}G \omega^5 L^{k} J^{lk}(\omega) \,.
\ee
These two Ward identity violations are resolved in two different ways, as it
 will be shown in the next subsections.

\subsection{Amplitudes involving a quadratic-interaction vertex}
\begin{figure}
  \begin{center}
      \begin{tikzpicture}
      	\draw [black, thick] (0,0) -- (6,0);
      	\draw [black, thick] (0,-0.1) -- (6,-0.1);
      	\draw[decorate, decoration=snake, line width=1.5pt, green] (0.5,0)  arc (180:0:1.75);
      	\draw[decorate, decoration=snake, line width=1.5pt, green] (4,0) -- (6,2);
      	\filldraw[black] (0.5,-0.1) circle (3pt) node[anchor=north] {$I^R\,,J^R$};
      	\filldraw[black] (4.,-0.1) circle (3pt) node[anchor=north] {$E\,,L^i$};
     \draw[-stealth,thick] (1.75,1.25)  arc (150:30:.5);
    \end{tikzpicture}\hspace{2cm}
    \begin{tikzpicture}
      	\draw [black, thick] (0,0) -- (6,0);
      	\draw [black, thick] (0,-0.1) -- (6,-0.1);
      	\draw[decorate, decoration=snake, line width=1.5pt, green] (0.5,0)  arc (180:0:1.75);
      	\draw[decorate, decoration=snake, line width=1.5pt, green] (4,0) -- (6,2);
      	\filldraw[black] (0.5,-0.1) circle (3pt) node[anchor=north] {$E\,,L^i$};
      	\filldraw[black] (4.,-0.1) circle (3pt) node[anchor=north] {$I^R\,,J^R$};
     \draw[-stealth,thick] (1.75,1.25)  arc (150:30:.5);
    \end{tikzpicture}
    \caption{Emission amplitudes involving quadratic-interaction vertices. The arrow indicate the direction of the retarded propagator in the loop, as dictated by the in-in formalism. The processes described by the right diagram have vanishing amplitude because $E$ and $L^i$ are conserved quantities.}    
    \label{fig:Ade}
  \end{center}
\end{figure}
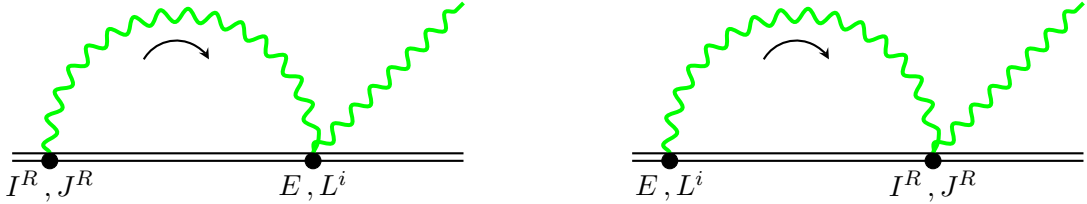

Processes like the ones shown in Fig.~\ref{fig:Ade} must be also considered, as
they are of the same $G$ order as the (f)tail ones.
It is straightforward to observe that such diagrams involve time derivatives of
the multipole linearly coupled to gravity, implying that the right diagram in
Fig.~ \ref{fig:Ade} is actually vanishing.

We then focus on the left diagram and first consider the multipole
to be $I^{ij}$. The angular momentum interaction vertex at quadratic order in
the gravitational field is \cite{Porto:2005ac}:
\be
S^{\rm spin}_{\rm source}
= - \int_{t} \frac12 J^{k|l} \left[ h_{0k,l} + \frac14 h_{l\lambda} \dot{h}_k^{\lambda} + \frac12 h_l^{\lambda} h_{0\lambda,k}
-\frac12 h_l^{\lambda}h_{0k,\lambda}+{\cal O}\pa{h^3}\right]\,,
\ee
where $\lambda$ is a space-time index, lowered and raised with $\eta_{\mu\nu}$.

A straightforward evaluation gives the following amplitude (here again $\K=\omega{\bf \hat n}$)
\be
\label{eq:aLtaile}
\ba{rcl}
i\ds{\mathcal{A}}^{(e-L-{\rm quad})}_{00}  (\omega,\omega{\bf \hat n})&=&\ds 0\,,\\
i\ds{\mathcal{A}}_{0l}^{(e-L-{\rm quad})}  (\omega,\omega{\bf \hat n})&=&\ds - 
i\frac{G}2 \omega^3 k_k L^j \epsilon^{ij[k} I^{l]i}(\omega)\,,\\
i\ds{\mathcal{A}}_{kl}^{(e-L-{\rm quad})} (\omega,\omega{\bf \hat n}) &=&\ds
 -i\frac{G}2\omega^4 L^j \epsilon^{ij(k}I^{l)i}(\omega)\,,
\ea
\ee
whose divergence is exactly opposite to the one reported in (\ref{eq:Weviol}).
By replacing $I_{ij}$ with any higher multipole of electric type, one obtains instead a vanishing result because of tracelessness of the multipoles themselves
(the amplitude can have at most two free indices and there are not enough
$k^j$ factors to contract with all the multipole indices).

It follows that the amplitude combination
$\mathcal{A}^{(e-L-\rm tot)}\equiv\mathcal{A}^{(e-L{\rm -ftail})}+\mathcal{A}^{(e-L-\rm quad)}$
satisfies the Ward identities $k^\mu\mathcal{A}^{(e-L-\rm tot)}_{\mu\nu}=0$ even for $r=0$,  and its expression is
\be
\label{eq:eLtot}
&&i\pa{\mathcal{A}^{(e-L{\rm -tot})}_{ij}}^{TT}(\omega,\omega {\bf \hat{n}}) = (-i)^r c_r^{(I)}G \Lambda^{TT}_{ij\,,kl}\epsilon_{mnq}\frac{i\omega^2L^q I^{pR(k}(\omega)}{(r+1)(r+2)(r+3)(r+4)}\nonumber\\
&&\quad\quad\quad\quad\quad\quad\times\bigg\{ k_n\pa{2[6+r(r+4)]\delta_{l)m} k_p k_R -r(10+r(r+5))\delta_{l)p}\delta_{i_1 m}k_{R-1}\omega^2}\bigg\}\,.
\ee

All the other cases, that is when $E$ and/or the magnetic multipoles $J^R$ are
involved, give vanishing results, again because virtue of the tracelessness of
the multipole moments.
To summarize, the class of processes represented in Fig.~\ref{fig:Ade} give a
  non-vanishing contribution only in the case they involve $L^i$ and $I^{ij}$,
which fixes the Ward identity violation of the corresponding $L$-ftail process.
  
\subsection{Ward-fixing amplitude correction}
\label{sec:Wfix}
The only Ward-violating amplitude left is the $L$-ftail involving the magnetic quadrupole.
Emitted waveforms which do not fulfill the Lorentz gauge condition have
already been treated in the standard multipolar PM formalism
\cite{Blanchet:1997ji,Blanchet:2013haa}.

The fix consists in finding a particular (non-unique)
solution $\bar{h}^{(W)}_{\mu\nu}$ of the homogeneous, linearized Einstein equations
\be
\Box \bar{h}^{(W)}_{\mu\nu} = 0\,,
\ee
whose divergence compensates the previously encountered Ward identity violation.
Once added to the previous anomalous solution $\bar{h}_{\mu\nu}$,
one will have
\be
\Box \pa{\bar{h}_{\mu\nu}+\bar{h}^{(W)}_{\mu\nu}} = \Lambda_{\mu\nu}\,,\quad  \partial^\mu\pa{\bar{h}_{\mu\nu}+\bar{h}^{(W)}_{\mu\nu}}=0\,. 
\ee

The tedious but straightforward check of the full, non-linearized Lorentz
condition is relegated to App. \ref{app:nlgauge}.

In the case we are interested in, the amplitude corresponding to the compensating terms $\bar{h}^{(W)}_{\mu\nu}$ is  
\be
\label{eq:aLtailm}
\ba{rcl}
\ds{\mathcal{A}}^{(W,m-L{\rm -ftail})}_{00} (\omega,\omega{\bf \hat n}) &=&\ds  \frac4{15} G \omega^3 k^l L^k J^{lk}(\omega)\,,\\
\ds{\mathcal{A}}_{0l}^{(W,m-L{\rm -ftail})}  (\omega,\omega{\bf \hat n})&=&\ds - \frac4{15}G \omega^4 L^k J^{lk}(\omega)\,,\\
\ds{\mathcal{A}}_{kl}^{(W,m-L{\rm -ftail})}  (\omega,\omega{\bf \hat n})&=&\ds 0 \,;
\ea
\ee
the $TT$ part of the amplitude in
Eq.~(\ref{eq:mLftail_amp}) is not affected by the anomaly fixing term
(\ref{eq:aLtailm}), whose space-space part is vanishing.
Interesting enough, if we did not include the quadratic-interaction amplitude and tried to fix also the electric quadrupole $L$-ftail following the multipolar PM formalism procedure,
we would have obtained a correcting amplitude exactly equal to eq.(\ref{eq:aLtaile}), as it is done in \cite{Henry:2023sdy}.

We have also checked that $\mathcal{A}^{(e-L-\rm tot)}$ and $\mathcal{A}^{(m-L-\rm tot)}\equiv\mathcal{A}^{(m-L{\rm -ftail})}$
return the correct contributions to the radiative multipole moments $U_R\,, V_R$, as they are defined and explicited in \cite{Faye:2014fra} for $r\leq 2$ and
for $r\leq 1$ in the electric and magnetic cases, respectively.

\section{Relation between self-energy diagrams and emission amplitudes}
\label{sec:cuts}

The self-energy diagrams can be factorized into products of emission diagrams or, equivalently, emission amplitudes can
be glued together to form self-energy amplitudes.
Note that self-energy amplitudes contribute to the dynamics of the
source \cite{Foffa:2019yfl}, hence, their
consistent computation is crucial to obtaining the correct effective Lagrangian
ruling the source dynamics.

For instance, the expression for the simplest self-energy diagram, as computed with the usual EFT Feynman rules \cite{Foffa:2019eeb}
\be
\label{eq:selfenergydiag}
i {\cal S}^{(I^2)} &=& \frac i{64\Lambda^2} \int \frac{{\rm d}\omega}{2\pi} \omega^4 I_{ij}(\omega)I_{kl}^{*}(\omega) \int_{\K} \frac{1}{\K^2-\omega^2}\times
\nonumber\\
&&\qquad \left[ \delta_{ik} \delta_{jl} + \delta_{il} \delta_{jk} - \frac{2}{(d-1)}\delta_{ij}\delta_{kl} + \frac{2}{(d-1)\omega^2} (k_i k_j \delta_{kl} + k_k k_l \delta_{ij})\right. \nonumber\\
&&\qquad\left.- \frac{1}{\omega^2} (k_i k_k \delta_{jl} + k_i k_l \delta_{jk} + k_j k_k \delta_{il} +k_j k_l \delta_{ik}) + \frac{4}{c_d \omega^4} k_i k_j k_k k_l \right]\,,
\ee
with $c_d\equiv 2(d-1)/(d-2)$, is identically equivalent to
\be
\label{eq:selfe}
i \tilde{\cal S}^{(I^2)} &=& - \frac{1}{2\Lambda^2} \int_{\K} \frac{d\omega}{2\pi}  \mathcal{A}^{(e,\, r=0)}_{\mu\nu}(\omega,\K){\cal P}[h_{\mu\nu},h_{\rho\sigma}]\mathcal{A}^{(e,\, r=0)}_{\rho\sigma}(-\omega,-\K)\nonumber\\
\label{eq:selfWard}
&=& - \frac{1}{2} \int_{\K} \frac{d\omega}{2\pi}  \pa{\mathcal{A}^{(e,\, r=0)}_{ij}}^{TT}(\omega,\K){\cal P}[h_{ij},h_{kl}]\pa{\mathcal{A}^{(e,\, r=0)}_{kl}}^{TT}(-\omega,-\K)\,,
\ee
with $\pa{\mathcal{A}^{(e)}_{ij}}^{TT}\equiv\Lambda^{TT}_{ij,ab}({\bf n})\mathcal{A}^{(e)}_{ab}$, and the standard in-out formalism with \emph{Feynman's} prescription
for the Green's functions is understood.
As a consequence of the Ward identities, the TT part alone of the emission
amplitude is sufficient to reconstruct the self-energy diagram.
The equality between $\cal S$ and $\tilde{\cal S}$ holds for all the electric
and magnetic multipoles, and it applies also to generic tail diagrams, see
App.~\ref{app:glue}).

Actually, this LO self-energy diagram is purely dissipative: after
$\K$-integration in (\ref{eq:selfenergydiag},\ref{eq:selfe}), one finds purely
imaginary $\omega$-integrand in ${\cal S}^{(I^2)}$.
Via standard optical theorem they can be related to a probability loss,
or, after multiplying the $\omega$-integrand by $|\omega|$, to energy emission.\footnote{When computing the self-energy diagram with Feynman Green's functions, the
$\omega$ integrand is complex and it is \emph{not} the Fourier transform of
a real function in direct space.}

Using more appropriately the
\emph{in-in} formalism, i.e., doubling the degrees of freedom and applying
retarded/advanced Green's functions
\cite{Galley:2009px,Schwinger:1960qe,Keldysh:1964ud}, one finds
the same real part of the self-energy action as with the in-out (after
appropriate identification of the doubled degrees of freedom, see
App.~\ref{app:eom-inin}) and the imaginary part of the $\omega$-integrand
is the Fourier transform of a real functional, which can be used to generate
non-conservative equations of motion.

In summary, as mentioned in the introduction and derived in
\cite{Foffa:2021pkg}, as far as self-energy diagrams with \emph{only two}
radiative Green's functions are concerned, one can consistently use the in-out
formalism with Feynman Green's functions to obtain the conservative part of the
equations of motion, and the optical theorem for the emitted energy.
The in-in formalism is, however, necessary to derive directly the dissipative
part of the equations of motion, see App. \ref{app:eom-inin} for details.

One can verify that analogous relations hold for the $M$-tail process,
\be
    i {\cal S}^{(e-M{\rm -tail})} &=& - G^2 E \frac{2^{r+2}(r+3)(r+4)}{(r+1)(r+2)(2r+5)!} \times \int \frac{d\omega}{2\pi} (\omega^2)^{r+3}I^{ijR}(\omega) I^{ij R}(-\omega) \left[ \frac{1}{\tilde{\varepsilon}} - \gamma_r^{(e)}\right]\,,\nonumber\\
\label{eq:SMtailse}
    &=& - \frac 1{2\Lambda^2} \int _{\K} \frac{{\rm d}\omega}{2\pi}\pa{\mathcal{A}^{e-M\rm-tail}_{ij}}^{TT}(\omega,\K)P[h_{ij},h_{kl}]\pa{\mathcal{A}^{(e)}_{kl}}^{TT}(-\omega,-\K)\,,\\
    i {\cal S}^{(m-M{\rm -tail})} &=& - G^2 E \frac{2^{r+2}(r+2)^2(r+4)(r!)^2}{(2r+1)(2r+3)(2r+5)(2r)![(r+3)!]^2} \nonumber\\
&&\quad\times \int \frac{{\rm d}\omega}{2\pi} 
J^{b|iRj}(\omega) J^{b'|kR'l}(-\omega) \omega^{2r+6} [\delta_{bb'}\delta_{ik} + (r+1) \delta_{ib'}\delta_{kb}] \left[
\frac{1}{\tilde{\epsilon}} - \gamma^{(m)}_r\right]\,,\nonumber\\
\label{eq:SMtailsm}
&=& - \frac 1{2\Lambda^2} \int _{\K} \frac{{\rm d}\omega}{2\pi}\pa{\mathcal{A}^{m-M\rm-tail}_{ij}}^{TT}(\omega,\K)P[h_{ij},h_{kl}]\pa{\mathcal{A}^{(m)}_{kl}}^{TT}(-\omega,-\K)\,,
\ee
with
\be
\ba{rcl}
\ds\frac{1}{\tilde{\varepsilon}} &\equiv&\ds \frac{1}{d-3}+\log x- i \pi {\rm sgn}(\omega)\,,\\
\ds \gamma_{r}^{(e)} &\equiv&\ds \frac 12 (H_{r+\frac 52} - H_{\frac 12} + 2H_r +1) + \frac{2}{(r+2)(r+3)}\,,\\
\ds\gamma^{(m)}_r &=&\ds \frac{2}{r+3} + \frac{1}{2r+5} - \frac{1}{r+2} - \frac{1}{r+4} + H_{r+1} + \frac12 H_{r+\frac 32} + \log 2\,,
\ea
\ee
which are the same numbers, although written in different form, obtained
in \cite{Almeida:2021xwn} via direct computation of the self-energy diagrams.
Incidentally, this allows one to derive explicit relations between the
$\gamma_{r}^{(e,m)}$ coefficients appearing in (\ref{eq:SMtailse},\ref{eq:SMtailsm}) and the finite terms in the emission amplitudes, $\kappa_{r+2}\,,\pi_{r+2}$ of
(\ref{eq:Mtaile},\ref{eq:Mtailm}):
\be
\gamma_{r}^{(e)}&=& \ds\kappa_{r+2} - \left(\frac12 +\frac{1}{r+3} +\frac{1}{r+4} -\frac12 H_{r+\frac{5}{2}} - \log 2\right) \,,\\
\gamma_{r}^{(m)}&=& \ds\pi_{r+2} - \left(\frac12 +\frac{1}{r+3} +\frac{1}{r+4} -\frac12 H_{r+\frac{5}{2}} - \log 2\right)  + \frac{r+5}{2(r+3)} \,.
\ee

For the $L$-ftails we find
\footnote{Notice that the term carrying the $\delta_{0r}$ in the magnetic $TT$ emission amplitude happens to give a vanishing contribution to the self-energy.}:
\be
\label{eq:Ltail_se}
i {\cal S}^{(e-L\rm-tot)}&=&- G^2\frac{(12+50r +35 r^2 + 10 r^3 + r^4)}{(r+1)^2(r+2)^2(r+3)!(2r+5)!!} \epsilon^{ikl}L^l\ \int \frac{{\rm d}\omega}{2\pi}\, I^{ijR}(\omega)I^{kjR}(-\omega) \omega^{7+2r}\nn\\
   &=& - \frac 1{2\Lambda^2} \int _{\K} \frac{{\rm d}\omega}{2\pi}\pa{\mathcal{A}^{e-L\rm-tot}_{ij}}^{TT}(\omega,\K)P[h_{ij},h_{kl}]\pa{\mathcal{A}^{(e)}_{kl}}^{TT}(-\omega,-\K)\,,\\
\label{eq:Ltail_sm}
i {\cal S}^{(m-L\rm -tot)}&=&-G^2\frac{4 (36 +50 r+35 r^2+10 r^3+r^4)}{(r+1)^2 (r+3)^2(r+3)!(2r+5)!!}\epsilon^{ikl}L^l\ \int \frac{{\rm d}\omega}{2\pi}\, J^{ijR}(\omega)J^{kjR}(-\omega) \omega^{7+2r}\nn\\
&=& - \frac 1{2\Lambda^2} \int _{\K} \frac{{\rm d}\omega}{2\pi}\pa{\mathcal{A}^{m-L\rm-tot}_{ij}}^{TT}(\omega,\K)P[h_{ij},h_{kl}]\pa{\mathcal{A}^{(m)}_{kl}}^{TT}(-\omega,-\K)\,.
\ee
In the electric $r=0$ case one has
\be
\label{eq:ampSLQQ}
i {\cal S}^{(LI^2)}&\equiv&\left.i {\cal S}^{(e-L\rm-tot)}\right|_{r=0}= - \frac1{30}G^2 \epsilon^{ikl}L^l  \int \frac{{\rm d}\omega}{2\pi}\, I^{ij}(\omega)I^{jk}(-\omega) \omega^7\,,
\ee
also in agreement with \cite{Henry:2023sdy}, once the corresponding quadratic-interaction process is also added in the self-energy calculation.
Note that the numerical factor $1/30$ corrects the value $8/15$ obtained
via the incomplete computation in \cite{Foffa:2019eeb} which did not take into account
the process with quadratic interaction in Fig.~\ref{fig:Ade}.

One can check the result (\ref{eq:ampSLQQ}) by computing the radiated energy and angular momentum calculable from the $L$-ftail
quadrupolar emission amplitude corrected by the quadratic-interaction process,
and comparing the result with the mechanical
energy and angular momentum loss derivable from the equations of motion generated by the
functional ${\cal S}^{(LI^2)}$, which are expected to agree with the former
modulo total derivatives, or \emph{Schott} terms \cite{schott,Bini:2012ji}.
It turns out that the contribution to the energy emission is a total
derivative, thus not being useful for our purposes; we then focus on angular momentum emission.

Starting from standard textbook formula, see e.g. eq.~(2.61) of
\cite{maggiore:gw1}, for the \emph{emitted} angular momentum
\be
\label{eq:Ldot}
\epsilon_{ijq}\dot{L}^q=\frac{r^2}{32\pi G}\int d\Omega\langle
\dot h^{TT}_{kl}x_i\partial_jh^{TT}_{kl}-2\dot h^{TT}_{kj} h^{TT}_{ki}
\rangle - i\leftrightarrow j\,,
\ee
and using the standard quadrupole formula for GWs one obtains the
leading order (LO) term
\be
\label{eq:Lflux}
\epsilon_{ijq}\left.\dot{L}^q\right|_{LO}=\frac {2G}5\pa{\langle \dddot I_{ik}\ddot I_{jk}\rangle-\langle\dddot I_{jk}\ddot I_{ik}\rangle}\,,
\ee
which matches the mechanical angular momentum loss obtained using the
Burke-Thorne acceleration $a_i^{(BT-I)}=-\frac{2G}5 x^jI^{(5)}_{ij}$
\cite{Burke:1970dnm}, modulo Schott terms.

Using the emission amplitude one has
\be
\pa{h_{ij}^{(e-L-\rm tot)}}^{TT}=\frac{2G} r\Lambda_{ij,kl}
\pa{\ddot I_{kl}-G\ddddot I_{a(k}\epsilon_{l)bq}L^q\hat n^a\hat n^b}\,,
\ee
which plugged into (\ref{eq:Ldot}) enables us to compute its contribution to
the angular momentum emission rate
\be
\label{eq:dL_Ltail_flux}
\epsilon_{ijq}\left.\dot{L}^q\right|_{LI^2}=
\frac{2G^2}{15}L^qI_{jk}^{(3)}I^{(4)}_{kl}\epsilon_{ilq}-i\leftrightarrow j\,,
\ee
which matches, again modulo Schott terms, the mechanical angular momentum loss
obtained from the modified Burke-Thorne acceleration
\be
\label{eq:BTLQ}
a_i^{(BT-LI)}=\frac{2G^2}{15}L^q
\pa{x^jI_{jk}^{(7)}\epsilon_{ikq}- x^jI^{(7)}_{ik}\epsilon_{jkq}}\,.
\ee
Acceleration (\ref{eq:BTLQ}) can be obtained from the in-in version of the
effective action (\ref{eq:ampSLQQ}), see App. \ref{app:eom-inin} for details.
Note that (\ref{eq:dL_Ltail_flux}) agrees also with standard results, see
e.g.~eq.~(2.7) of \cite{Arun:2009mc}, giving further confirmation that the
$1/30$ coefficient in the expression for ${\cal S}^{(LI^2)}$ is indeed
correct.

Eq.~(\ref{eq:dL_Ltail_flux}) is the leading order term
of a series of angular momentum flux contributions by $L$-ftails
involving electric and magnetic multipoles of all orders, which can be
straightforwardly derived either from the corresponding emission amplitudes
(\ref{eq:eLtot},\ref{eq:mLftail_amp}), or from the effective actions (\ref{eq:Ltail_se},\ref{eq:Ltail_sm}).

\section{Conclusion and discussion}
\label{sec:concl}

We have analyzed next-to-leading order far-zone diagrams contributing
to both conservative and dissipative two-body dynamics.
The interaction studied are of the tail-type, i.e., due to emission
of radiation which subsequently interacts with the quasi-static curvature generated by the mass and angular momentum of the source.

We applied field theory methods within the framework of NRGR
which makes use of standard gauge-fixed path integral formulation
to derive the classical effective action.
In analogy with what has been found in the classical approach of the multipolar post-Minkowskian
formalism, we found that, in some cases, the gauge condition chosen to make
the kinetic term of the gravitational field invertible is not respected by
loop corrected classical solutions, giving origin to anomalous scattering amplitudes.
We have then shown how such apparent anomalies are canceled, respectively in the electric and magnetic case, by the inclusion of a quadratic-interaction process,
and by the standard multipolar PM correction procedure.

Then, we showed the consequence of fixing Ward identity in emission diagrams
  for self-energy ones, by suitably obtaining the latter by glueing the former,
  i.e.~via \emph{generalized unitarity}.
As a natural prosecution of the present work, we plan to study in an analogous
approach the memory process, which involves the emission of radiation
scattering off another radiative mode, which has both analogies
and differences with respect to the tail and tail-like processes studied
in the present work.

While a violation and the recovery of the gauge-fixing condition is also
expected in the memory case, as per the results of \cite{Blanchet:1997ji}, the
presence of three radiative degrees of freedom in the memory self-energy
amplitude requires a thorough treatment within the in-in formalism which will be
the subject of a future investigation, together with the fundamental origin of the violation of the energy momentum conservation.

\section*{Acknowledgments}

The authors thank Quentin Henry and Fran\c{c}ois Larrouturou for useful discussions and the FAPESP grant 2021/14335-0 as part of this work was done during the
program ``Gravitational Waves meet Amplitudes in the Southern Hemisphere''.
R.S. acknowledges support by CNPq under grant n. 310165/2021-0 and
by FAPESP grant n.2022/06350-2.
S.F. is supported by the Fonds National Suisse, grant $200020\_191957$, and by the SwissMap National Center for Competence in Research. A.M. also acknowledges support by CNPq, under grant 163090/2022-0.
The work of G.L.A. is financed in part by the Coordenação de Aperfeiçoamento de Pessoal de Nível Superior—Brasil (CAPES)—Finance Code 001.

\appendix

\section{Un-integrated tail emission amplitudes}
\label{app:MLtailui}

\be
 i{\cal A}^{(e-M\rm-tail)}_{\mu\nu} (\omega,\K)= \frac{(-i)^{r+1} c_r^{(I)}}{\Lambda^2} E I^{ijR}(\omega) \ds\int_\Q \frac{(k+q)_{R}}{[(\K+\Q)^2-\omega^2]\Q^2}f^{(e-M)}_{\mu\nu}\,,\\
 i{\cal A}^{(m-M\rm-tail)}_{\mu\nu} (\omega,\K)= \frac{(-i)^{r+1} c_r^{(J)}}{\Lambda^2} E J^{b|iRa}(\omega) \ds\int_\Q \frac{(k+q)_b(k+q)_{R}}{[(\K+\Q)^2-\omega^2]\Q^2}f^{(m-M)}_{\mu\nu}\,.
\ee
\be
f^{(e-M)}_{00}&=&\left(-\frac{i}{4}\right)\left[ \frac{1}{c_d} (k+q)_i (k+q)_j (\K+\Q)\cdot\Q + (k_i k_j+k_i q_j+q_i q_j)\omega^2\right]\,,\\
f^{(e-M)}_{0k}&=&\left(\frac{i\omega}{4}\right) 
\left\{\frac{1}{c_d} (k+q)_i (k+q)_j q_k + (k+q)_i \left[ (\K+\Q)\cdot\K\delta_{jk} - k_j q_k + k_k q_j\right] - \omega^2\delta_{jk} q_i\right\}\,,\nonumber\\
&&\ \\
f^{(e-M)}_{kl}&=&\left(-\frac{i}{4}\right) 
\bigg\{  \omega^4\delta_{ik}\delta_{jl} + \omega^2 (k+q)_i \left[ q_j\delta_{kl} - (q_k\delta_{jl}+q_l\delta_{jk})\right] \nonumber\\
&&\qquad\qquad\qquad+ \frac{1}{c_d} (k+q)_i (k+q)_j \left[ k_k q_l + k_l q_k + 2q_k q_l - (\K+\Q)\cdot\Q \delta_{kl}\right]
\bigg\}\,,\\
f^{(m-M)}_{00}&=&\left(-\frac18\right)\omega (k_a q_i + k_i q_a)\,,\\
f^{(m-M)}_{0k}&=&\left(\frac18\right)\left[ \omega^2 (q_i \delta_{ak} + q_a \delta_{ik}) - (k+q)_i (k+q)_c (k_c \delta_{ak} - k_a \delta_{ck})
\right]\,,\\
f^{(m-M)}_{kl}&=&\left(\frac18\right) \omega \left[\omega^2 (\delta_{ak} \delta_{il} + \delta_{ik} \delta_{al}) - (k+q)_i (q_l \delta_{ak} + q_k \delta_{al}) + \delta_{kl} (k+q)_i q_a \right]\,,
\ee
\be
 i{\cal A}^{(e-L\rm-tail)}_{\mu\nu} (\omega,\K)= \frac{(-i)^{r} c_r^{(I)}}{\Lambda^2} J^{b|a} I^{ijR}(\omega) \ds\int_\Q \frac{q_a (k+q)_{R}}{[(\K+\Q)^2-\omega^2]\Q^2}f^{(e-L)}_{\mu\nu}\,,\\
 i{\cal A}^{(m-L\rm-tail)}_{\mu\nu} (\omega,\K)= \frac{(-i)^{r} c_r^{(J)}}{\Lambda^2} J^{s|t} J^{b|iRa}(\omega) \ds\int_\Q \frac{q_t(k+q)_{b'}(k+q)_{R}}{[(\K+\Q)^2-\omega^2]\Q^2}f^{(m-L)}_{\mu\nu}\,,
\ee
\be
f^{(e-L)}_{00}&=& \left( -\frac{i\omega}{8}\right)  \left\{ \delta_{jb} \omega^2 q_i + (k+q)_j \left[k_b(2k_i-q_i) + 3(\K+\Q)\cdot\Q \delta_{ib}\right] \right\}\,,\\
f^{(e-L)}_{0k}&=&\left( \frac{i}{8}\right)  \left\{ 
(k+q)_i (k+q)_j (\K\cdot\Q \delta_{bk}-k_b q_k)  \right.\nonumber\\
&&\qquad\qquad\left.+ [\K\cdot\Q \delta_{ib}\delta_{jk} - \delta_{bk} q_i q_j + q_j(k+q)_k \delta_{ib} + (2k+q)_j k_b \delta_{ik}] \omega^2\right\}\,,\nonumber\\
&&\ \\
f^{(e-L)}_{kl}&=&\left( \frac{i\omega}{16}\right) (\delta_{kc}\delta_{ld}+\delta_{kd}\delta_{lc}) \left\{ - \Big[ q_j \delta_{ib}\delta_{cd} + 2(q_i \delta_{bc} - q_c \delta_{ib} + k_b \delta_{ic}) \delta_{jd} \Big] \omega^2 \right.
\nonumber\\
&&\qquad+ (k+q)_j \Big[ 2(k+q)_c q_i \delta_{bd}  - k_b q_i \delta_{cd} -2(k+q)_c q_d \delta_{ib} + (\K+\Q)\cdot\Q \delta_{cd} \delta_{ib}\nonumber\\
&&\qquad\qquad+\left.+ 2k_b q_d \delta_{ic} -2(\K+\Q)\cdot\Q \delta_{bd} \delta_{ic} \Big]\right\}\,,\nonumber\\
f^{(m-L)}_{00}&=&\left(-\frac1{16}\right) \left[ \omega^2 (q_i \delta_{as} - k_a \delta_{is}) + (k+q)_i \left( k_a (3k+2q)_s + 3 (\K+\Q)\cdot\Q \delta_{as}\right)\right]\,,\\
f^{(m-L)}_{0k}&=&\left(\frac\omega{16} \right) \left[
(2k+q)_i k_s \delta_{ak} +(k+q)_k q_i \delta_{as} + (k_a k_s + \K\cdot\Q \delta_{as}) \delta_{ik} \right.\\
&&\qquad\qquad\qquad\qquad\qquad\left.+ (\K\cdot\Q \delta_{ak}-k_a(k+q)_k) \delta_{is} + 2k_a q_i \delta_{ks} \right]\,,\nonumber\\
f^{(m-L)}_{kl}&=&\left(\frac1{16}\right)\bigg\{
(k+q)_i \Big[ (k_k q_l+k_l q_k) \delta_{as} + 2 q_k q_l \delta_{as} + k_a (k+q)_l \delta_{sk}\nonumber\\
&&\qquad\qquad\qquad+ (\K+\Q)\cdot\Q (\delta_{al}\delta_{sk} + \delta_{ak}\delta_{sl})+ k_a (k+q)_k \delta_{sl} - k_s (q_k \delta_{al} + q_k \delta_{al}) \Big]  \nonumber\\
&&\qquad- \omega^2 \Big[
(q_k \delta_{il} + q_l \delta_{ik}) \delta_{as} + (q_k \delta_{al} + q_l \delta_{ak}) \delta_{is} -q_i (\delta_{ak}\delta_{sl}+\delta_{al}\delta_{sk}) \Big]  \nonumber\\
&&\qquad +\delta_{kl}\Big[\omega^2 (q_i \delta_{as} - k_a \delta_{is}) - k_i \delta_{as} (\K+\Q)\cdot\Q - q_i (\K+\Q)\cdot\Q \delta_{as} - (k+q)_i k_a k_s\Big]\nonumber\\
&&\qquad+ k_a (\delta_{ik}\delta_{sl}+\delta_{il}\delta_{sk})-2 k_s (\delta_{al}\delta_{ik}+\delta_{ak}\delta_{il})  \bigg\}\,.				
\ee

\section{Mechanical angular momentum flux and in-in formalism}
\label{app:eom-inin}

The real parts of the effective actions associated to tail and ftail processes via
Feynman Green's functions carry information about the time-symmetric
part of the equations of motion.
Their imaginary parts carry the information of the probability loss,
from which it is possible to recover the energy and angular momentum
loss with standard methods \cite{Sturani:2021ucg,Goldberger:2022rqf}.

However, adopting the in-in formalism \cite{Schwinger:1960qe,Keldysh:1964ud},
it is possible to derive time-asymmetric equations of motion by writing
a functional ${\cal S}_{in-in}$
for a degree of freedom propagating forward in time ``1'' and one backwards
``2'':
\be
{\cal S}_{\rm in-in}=\int dt\,\pa{{\cal L}_1-{\cal L}_2}\,.
\ee
The in-in functional is then written in terms of the Keldysh $+,-$ variables
defined, for a generic dynamical variable $x$, as
$x_+\equiv (x_1+x_2)/2$, $x_-\equiv x_1-x_2$, in terms of which e.g.~\eqref{eq:ampSLQQ} can be recast into its in-in counterpart
\be
   {\cal S}^{(LI^2)}_{\rm in-in}=\frac{G^2}{30}\int{\rm d}t
   \pa{\ddddot I_{+ik} \dddot I_{-jk}+\ddddot I_{-ik} \dddot I_{+jk}}\epsilon_{ijq}L^q\,,
   \ee
   and the equation of motion (\ref{eq:BTLQ}) can be derived from
   $\frac{\delta S^{(LI^2)}_{in-in}}{\delta x_-}|_{x_-=0}=0$, and by taking the physical limit $x_+ \rightarrow x$, leading to
\be
\epsilon_{ijq}\left.\dot{L}^q\right|_{(BT-LI^2)}=
-\frac{2G^2}{15}L^qI_{jk}\pa{I^{(7)}_{kl}\epsilon_{ilq}-I^{(7)}_{il}\epsilon_{klq}}
-i\leftrightarrow j\,.
\ee
The angular momentum loss involving the magnetic quadrupole $L$-ftail is derived along the same lines.

\section{Non-linearities in the gauge conditions}
\label{app:nlgauge}

The gauge we have been using to compute amplitudes is the harmonic gauge, defined by
\begin{equation}\label{harmgauge}
\Gamma^\alpha_{\mu\nu} g^{\mu\nu} = 0\,.
\end{equation}
When expanded to first order in the fields, we obtain the condition
\begin{equation}
\partial^\mu \bar{h}_{\mu\nu} = \partial^\mu \left( h_{\mu\nu} - \frac12 \eta_{\mu\nu} h \right) = 0\,.
\end{equation}
This condition holds only for the leading-order processes. For higher orders, on the other hand, like the tails and $L$-ftails studied in this paper, we have to solve Eq.~\eqref{harmgauge} iteratively in $G$.
In this case, the general structure of the problem can be organized as
\begin{equation}\label{h1h2h3}
\begin{aligned}
\partial^\mu \bar{h}^{(1)}_{\mu\nu} &= 0\,,\\
\partial^\mu \bar{h}^{(2)}_{\mu\nu} &= \lambda^{(2)}_{\mu\nu}(h^{(1)},h^{(1)})\,,\\
\partial^\mu \bar{h}^{(3)}_{\mu\nu} &= \lambda^{(3)}_{\mu\nu}(h^{(1)},h^{(1)},h^{(1)}) + \gamma^{(3)}_{\mu\nu}(h^{(2)},h^{(1)})\,,\\
&\cdots\,,
\end{aligned}
\end{equation}
where $\bar{h}^{(n)}_{\mu\nu}$ represents processes of order $G^n$ and $\lambda^{(n)}_{\mu\nu}$, $\gamma^{(n)}_{\mu\nu}$, etc., are functions of the lower-order perturbations $h^{(n-1)}, h^{(n-2)},\dots,h^{(1)}$. In particular, at order $G^2$, we have
\begin{equation}\label{gaugecondcorrect}
\partial^\mu \bar{h}^{(2)}_{\mu\alpha} = h^{(1)\mu\nu} \left( h^{(1)}_{\alpha\mu,\nu} -\frac12 h^{(1)}_{\mu\nu,\alpha}\right)\,.
\end{equation}

Below we show that, for processes of order $G^2$, the functions appearing on the right-hand side of eqs.~\eqref{h1h2h3} vanish on shell, and therefore, do not contribute for self-energy diagrams of tail-like processes. To show this, consider two processes of order $G$,  given generically by
\begin{equation}
h^{(1)}_{\mu\nu} = G \int\frac{d\omega}{2\pi} A^{R}(\omega) \int_\K \frac{e^{-i\omega t+ i \K\cdot\X}}{\K^2-\omega^2} K_L\,,
\quad \text{and} \quad
h^{'(1)}_{\mu\nu} = G \int\frac{d\omega'}{2\pi} B^{R'}(\omega') \int_\Q \frac{e^{-i\omega' t+ i \Q\cdot\X}}{\Q^2-\omega'^2} Q_{L'}\,,
\end{equation} 
where $A^{R}(\omega)$ and $B^{R'}(\omega')$ represent the integrand for arbitrary multipoles, including the ones related to conserved multipoles, by making, e.g., $A^{R}(\omega) \rightarrow E \delta(\omega)$.
$K_L$ represents any combination of the momenta $k$'s, and likewise for $Q_{L'}$. By plugging this into the right-hand side of Eq.~\eqref{gaugecondcorrect}, we encounter the following behavior:
\begin{align}
\partial^\mu \bar{h}_{\mu\nu} &\sim G^2 \partial^\mu \left[ \int\frac{d\omega}{2\pi} A^{R}(\omega) \int_\K \frac{e^{-i\omega t+ i \K\cdot\X}}{\K^2-\omega^2} K_L
\times
\int\frac{d\omega'}{2\pi} B^{R'}(-\omega') \int_\Q \frac{e^{i\omega' t- i \Q\cdot\X}}{\Q^2-\omega^2} Q_{L'} \right] \nonumber\\
&\rightarrow \int_\K\frac{d\omega}{2\pi}  \frac{e^{-i \omega t+i \K\cdot\X} }{\K^2-(\omega+i{\tt a})^2}
 \times \frac{\left[  k^\mu \bar{\mathcal{A}}_{\mu\nu}
 \right]}{\Lambda^2}\,, 
\end{align}
where
\begin{equation}
k^\mu \bar{\mathcal{A}}_{\mu\nu} = (\omega^2-\K^2)
 \int\frac{d\omega'}{2\pi} A^{R}(\omega+\omega')  B^{R'}(-\omega')\int_{\Q} \frac{(KQ)_{LL'}}{[(\K + \Q)^2-(\omega+\omega')^2](\Q^2-\omega'^2)}\,.
\end{equation}
Notice that this expression is always vanishing on-shell, and hence, will not
play any role in the construction of self-energy diagrams for tail-like processes, see the appendix below.
This justifies the use of the linearized Lorentz condition
$\partial^\mu\bar{h}_{\mu\nu}=0$ for processes of order $G^2$.

\section{Cutting and gluing amplitudes}
\label{app:glue}
We present in this section a heuristic derivation of generalized unitarity
  applied to
tail-like self-energy diagrams.
The gluing of emission amplitudes can be written as
\be
\label{eq:amp_glue}
\ba{rcl}
\ds i\tilde{\cal S}^{\rm(tail)}&=&\ds -\frac{i}{\Lambda^2}\int_{\K}\frac{d\omega}{2\pi}
\frac{{\mathcal{A}^{\rm(tail)}_{ij}}^{TT}(\omega,\K){{\cal A}^{(LO)}_{ij}}^{TT}(-\omega,-\K)}{\K^2-\omega^2}\\
&=&\ds -\frac{i}{16\pi^2\Lambda^2}\int{\rm d}\Omega\int\frac{d\omega}{2\pi}
(i\omega) {{\cal A}^{\rm(tail)}_{ij}}^{TT}(\omega,\omega{\bf\hat n})
{{\cal A}^{(LO)}_{ij}}^{TT}(-\omega,-\omega{\bf\hat n})\\
&=&\ds -\frac{i}{16\pi^2\Lambda^2}\int_t\int{\rm d}\Omega
\dot{\cal A}^{\rm (tail)\,TT}_{ij}(t,{\bf \hat n})
{{\cal A}^{(LO)}_{ij}}^{TT}(t,{\bf \hat n})\,,
\ea
\ee
the first passage holding because of the useful identity:
\be
\label{eq:intk_polew}
\int_{\K}\frac{k_{i_1}\dots k_{i_{2l}}}{\K^2 -(\omega\pm i\tt{a})^2}=
\pa{ \pm i\frac{\omega}{4\pi}}\delta_{i_1\dots i_{2l}}\frac{\omega^{2l}}{(2l+1)!!}=
\pa{ \pm i\frac{\omega}{16\pi^2}}\omega^{2l}\int{\rm d}\Omega\,\hat n_{i_1}\dots \hat n_{i_{2l}}\,,
\ee
which can be inserted into (\ref{eq:amp_glue}) as ${\cal A}^{\rm(tail)},{\cal A}^{(LO)}$ have no
poles in $\K$.
To derive the non time-symmetric equations of motion, it is necessary to
recast the action (\ref{eq:amp_glue}) into its in-in counterpart, as
described in App.\ref{app:eom-inin}, to obtain the mechanical energy loss
\be
\label{eq:av}
-{\bf a}\cdot{\bf v}=\frac{1}{16\pi^2\Lambda^2}\int {\rm d}\Omega
\dddot{\cal{A}}^{\rm(tail)\,TT}_{ij}(t,{\bf\hat n})
\pa{{{\cal A}^{(LO)}_{ij}}^{TT}(t,{\bf\hat n})}^{(-1)}\,,
\ee
where a double integration by parts over ${\cal A}^{(LO)}$, which is linear in
$\ddot{I}_{ij}$, introduces the ``antiderivative''
$\pa{{\cal A}^{(LO)}_{ij}}^{(-1)}\propto \dot{I}_{ij}$.

Had we used the Feynman boundary condition in (\ref{eq:intk_polew}), we would have
obtained $-i|\omega|$ instead of $\mp i\omega$ in the first parentheses,
giving rise to the standard optical theorem relationship between self-energy
imaginary part and emission probability, which can be related to the energy loss
by multiplying the $\omega$-integrand by $|\omega|$.
While a consistent use of retarded/advanced Green's functions requires the
in-in formalism, see App.\ref{app:eom-inin} \cite{Galley:2015kus},
if we limit ourselves to the conservative dynamics and the computation of the
energy flux one can use Feynman Green's functions.

The energy loss of eq.~(\ref{eq:av}) agrees, modulo Schott terms,
with the one obtained by direct computation of the gravitational luminosity at
infinity ${\cal F}$, via the asymptotic GW waveform
\be
h_{ij}^{TT}&\simeq&-\frac1{4 \pi r}\int_\omega \pa{-\frac1{\Lambda^2}}  {\rm e}^{-i \omega t_{\rm ret}} \dot{\mathcal{A}}^{TT}
_{ij}(\omega,{\bf n}\omega)=\frac1{4 \pi r \Lambda^2} \dot{\mathcal{A}}^{TT}_{ij}({\bf n},t_{\rm ret})\,,
\ee
as 
\be
\label{eq:fluxTT}
      {\cal F}=\Lambda^2 r^2\int {\rm d}\Omega\ \dot{h}_{ij}^{TT}\dot{h}_{ij}^{TT}=
      \frac1{16\pi^2\Lambda^2}\int {\rm d}\Omega\,
      \dot{\cal A}^{TT}_{ij}(t_{\rm ret},{\bf \hat n})
      \dot{\cal A}^{TT}_{ij}(t_{\rm ret},{\bf \hat n})\,,
\ee
with
\be
{\cal A}^{TT}_{ij}\simeq{{\cal A}^{(LO)}_{ij}}^{TT}+{{\cal A}^{\rm(tail)}_{ij}}^{TT}\,,
\ee
and expanding at NLO. 

Note that, considering the extra $i$ provided by the integration over $\K$, see
eq.~(\ref{eq:intk_polew}),
one has that the real part of ${\cal A}^{\rm(tail)}/{\cal A}^{(LO)}$ contributes
to the probability and energy loss, the imaginary part to the conservative
dynamics\footnote{Note that the Fourier transform of a direct space real
function is in general complex, here by real part of the $\omega$
integrand $A(\omega)$ of the effective action we mean a function satisfying
$A^*(\omega)=A(-\omega)$.},
and the self-energy action is completely determined by the emission amplitude.

\bibliography{../../../bibliography}

\end{document}